\documentclass[11pt]{article}

\usepackage[utf8]{inputenc}
\usepackage[T1]{fontenc}
\usepackage[margin=1in]{geometry}
\usepackage{url,hyperref,lineno,microtype,subcaption,graphicx}
\usepackage{setspace}
\usepackage{titlesec}
\usepackage{authblk}
\usepackage{amsmath,amssymb}
\usepackage[round,authoryear]{natbib}

\titleformat{\section}
  {\normalfont\Large\bfseries\sffamily}
  {}{0pt}{\MakeUppercase}
\titlespacing*{\section}{0pt}{18pt}{12pt}

\titleformat{\subsection}
  {\normalfont\large\bfseries}
  {}{0pt}{}
\titlespacing*{\subsection}{0pt}{12pt}{8pt}

\titleformat{\subsubsection}
  {\normalfont\normalsize\itshape}
  {}{0pt}{}
\titlespacing*{\subsubsection}{0pt}{10pt}{6pt}

\title{\textbf{Robust Nasality Representation Learning for Cleft Palate-Related Velopharyngeal Dysfunction Screening in Real-World Settings}}

\author[1]{Weixin Liu}
\author[2]{Bowen Qu}
\author[3]{Amy Stone}
\author[3]{Maria E. Powell}
\author[4]{Shama Dufresne}
\author[5,6]{Stephane Braun}
\author[5,6]{Izabela Galdyn}
\author[5,6]{Michael Golinko}
\author[1,2,7,8]{Bradley Malin}
\author[1,2,7]{Zhijun Yin\textsuperscript{*}\thanks{Correspondence: zhijun.yin@vumc.org; matthew.e.pontell@vumc.org}}
\author[5,6]{Matthew E. Pontell\textsuperscript{*}}

\affil[1]{Department of Electrical and Computer Engineering, Vanderbilt University, Nashville, TN, United States}
\affil[2]{Department of Computer Science, Vanderbilt University, Nashville, TN, United States}
\affil[3]{Department of Otolaryngology---Head and Neck Surgery, Vanderbilt University Medical Center, Nashville, TN, United States}
\affil[4]{School of Medicine, Vanderbilt University, Nashville, TN, United States}
\affil[5]{Department of Plastic Surgery, Vanderbilt University Medical Center, Nashville, TN, United States}
\affil[6]{Division of Pediatric Plastic Surgery, Monroe Carell Jr. Children's Hospital, Nashville, TN, United States}
\affil[7]{Department of Biomedical Informatics, Vanderbilt University Medical Center, Nashville, TN, United States}
\affil[8]{Department of Biostatistics, Vanderbilt University Medical Center, Nashville, TN, United States}

\date{}

\begin{document}

\maketitle

\begin{abstract}
\noindent\textbf{Background:} Velopharyngeal dysfunction (VPD) is an impaired ability to achieve adequate velopharyngeal closure during speech, often resulting in hypernasality and reduced intelligibility. VPD screening and diagnosis require specialized expertise and controlled recording conditions, limiting scalable access outside high-income countries.

\noindent\textbf{Key challenge:} Speech-based machine learning models can perform extremely well under standardized clinical recording conditions. However, performance often deteriorates when deployed on consumer devices (e.g., phones or tablets) and in uncontrolled acoustic environments. This degradation is largely driven by \emph{domain shift} arising from differences in recording conditions (e.g., device and channel characteristics, background noise, and room acoustics), which can cause models to rely on spurious recording artifacts rather than pathology-relevant cues.

\noindent\textbf{Methods:} This study introduces a two-stage framework to improve robustness under realistic recording scenarios. \emph{Nasality representation pre-training} employs a nasality-focused representation via supervised contrastive learning (SupCon) using an auxiliary dataset with phoneme alignments to form oral-context versus nasal-context supervision. During \emph{Frozen-encoder VPD screening}, the encoder is frozen to perform VPD screening using lightweight classifiers on 0.5 second chunks with probability aggregation to produce recording-level decisions using a fixed decision threshold. Here, \emph{in-domain} refers to standardized clinical recordings used for model development, and \emph{out-of-domain} refers to heterogeneous public Internet recordings collected under uncontrolled conditions and evaluated without any adaptation. Data is then compared against prior-study baselines, including MFCC features and large pretrained speech representations, using the same evaluation protocol.

\noindent\textbf{Results:} On an in-domain clinical cohort of 82 subjects (60 train / 22 test; 345 training recordings; 131 test recordings; multiple recordings per subject), the proposed approach achieves perfect recording-level screening performance (macro-F1 = 1.000, accuracy = 1.000). On a separate out-of-domain set of 131 public Internet recordings, large pretrained speech representations degrade substantially, and MFCC is the strongest baseline (macro-F1 = 0.612, accuracy = 0.641). The proposed method achieves the best overall out-of-domain performance (macro-F1 = 0.679, accuracy = 0.695), improving over the strongest baseline by +0.067 macro-F1 and +0.054 accuracy (point-estimate improvements) under the same evaluation protocol and fixed threshold.

\noindent\textbf{Conclusion:} Learning a nasality-focused representation prior to clinical classification can reduce sensitivity to recording artifacts and improve robustness when moving from the laboratory to real-world audio recording scenarios. This design supports practical deployment of VPD screening and motivates domain-robust evaluation protocols for deployable speech-based digital health tools.

\vspace{0.5em}
\end{abstract}

\section{Introduction}

Velopharyngeal dysfunction (VPD) arises from inadequate function of the velopharyngeal port during speech, which allows for abnormal air coupling between the oral and nasal cavities. VPD frequently presents with hypernasality and impaired speech intelligibility, both of which negatively impact communication, feeding, and psychosocial function \citep{guyton2018acquired}. The most common manifestation of VPD is velopharyngeal insufficiency (VPI), which often results from velopharyngeal dysfunction in the setting of a cleft palate \citep{lucas2025machine}. In patients with cleft palate, VPI rates can exceed 30\% and are likely much higher worldwide \citep{alter2025s}. Diagnosis and management of VPD requires specialized speech-language pathologists (SLPs) that are often members of multidisciplinary cleft care teams, and evaluations are often conducted in standardized acoustic recording conditions. These resources are often not available in low- and middle-income countries (LMICs), thereby creating a barrier to timely diagnosis and management \citep{alter2025support}.

In an attempt to augment the reach of existing SLPs, teams have begun to explore the use of machine learning models to automatically detect the presence of VPD from acoustic samples \citep{lucas2025machine,shirk2025leveraging,alter2025support,liu2025out}. To ensure that models are clinically translatable, a scalable VPD screening tool must be deployable in the field and operate on consumer devices in everyday acoustic environments. However, clinical deployment introduces significant variation in acoustic setting, microphone frequency response, channel compression, background noise, reverberation, and speaking style, amongst many other factors. Such factors induce domain shift between standardized clinical recordings used for model development and real-world audio encountered during deployment \citep{liu2025out,ganin2016domain,sun2016deep}. A central challenge is that models trained in-domain can inadvertently focus on spurious correlations tied to recording conditions (e.g., device-specific spectral coloration or background noise patterns), achieving near-ceiling performance in controlled tests yet failing to generalize under realistic variability. This phenomenon is widely recognized as ``shortcut learning'' in medical AI, where models latch onto non-pathological artifacts rather than actual biological signals \citep{geirhos2020shortcut,degrave2021ai}.

Prior work on VPD speech analysis has explored both classical acoustic features and modern deep learning approaches, including objective assessment of hypernasality and nasality-related measures \citep{lucas2025machine,alter2025support,shirk2025leveraging,liu2025out,mathad2021deep,lozano2024computing,zhang2023automatic}. More recently, large self-supervised speech representations (e.g., Wav2Vec2, HuBERT, data2vec, Whisper encoders) have enabled strong performance on small clinical datasets by leveraging broad pretraining \citep{baevski2020wav2vec,hsu2021hubert,baevski2022data2vec,radford2023robust}. Despite these advances, high in-domain accuracy does not guarantee deployable robustness. Rather, many approaches are evaluated primarily within a single recording domain or under mild condition changes, leaving a gap in understanding and addressing generalization failures under substantial domain shift, for example, moving from the clinic to consumer devices \citep{liu2025out}.

\begin{figure}[htbp]
\centering
\IfFileExists{model_final_cut.png}{%
  \includegraphics[width=0.8\linewidth]{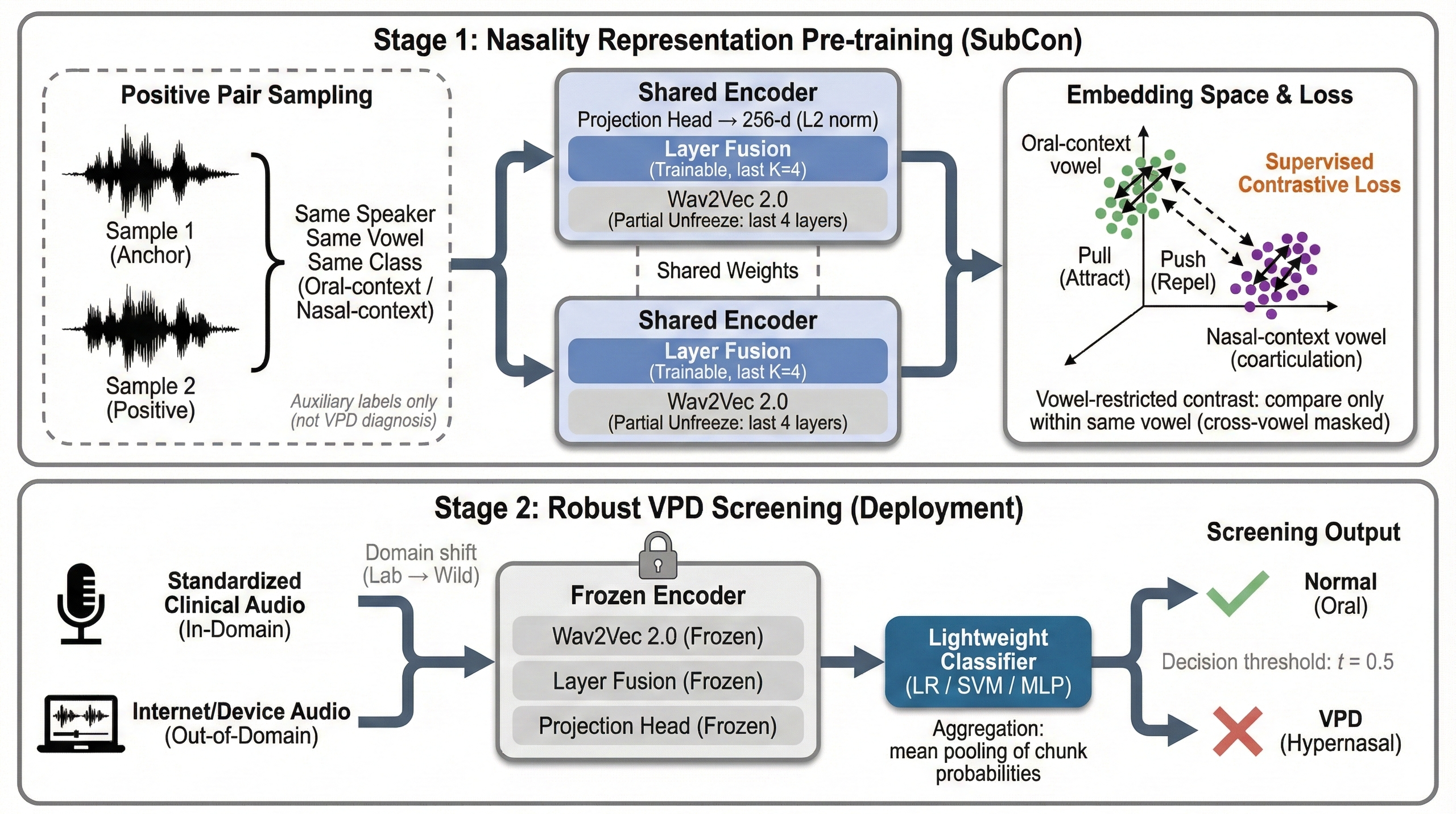}%
}{%
  \fbox{\parbox{0.95\linewidth}{\centering
  \vspace{1.5cm}
  Placeholder for Figure~\ref{fig:overview}\\
  (overview figure not included in this compile package)
  \vspace{1.5cm}}}%
}

\caption{Overview of the proposed two-stage framework. Stage 1: nasality representation pre-training using supervised contrastive learning (SupCon) with positive pairs sampled from the same speaker, same vowel, and same auxiliary class (oral-context vs.\ nasal-context), while restricting contrastive comparisons to within-vowel pairs to reduce phonetic-content leakage. A Wav2Vec2.0 backbone with trainable layer fusion and a projection head outputs 256-dimensional $\ell_2$-normalized embeddings. Stage 2: robust VPD screening under domain shift (lab $\rightarrow$ wild) using the frozen encoder as a feature extractor on 0.5\,s chunks, followed by a lightweight classifier (LR/SVM/MLP/XGBoost) and mean aggregation of chunk-level probabilities to produce recording-level screening decisions (in-domain and out-of-domain) using a fixed decision threshold.}
\label{fig:overview}
\end{figure}

This study aims to address this gap by learning a representation that targets the underlying production-related attribute, namely nasality, before training the clinical screening classifier, with the goal of reducing reliance on domain-dependent recording artifacts under cross-device and cross-environment variability \citep{liu2025out,ganin2016domain,sun2016deep}. Such nasality-related cues could be more stable across devices than many channel- or environment-specific artifacts, and thus a representation encouraged to encode nasality distinctions can improve robustness when the recording domain changes \citep{liu2025out}. To achieve this goal, we aim to use a two-stage framework (\textbf{Figure~\ref{fig:overview}}) to model nasality-related cues. With nasality representation pre-training, supervised contrastive learning (SupCon) is performed on an auxiliary speech corpus with phoneme alignments \citep{khosla2020supervised,panayotov2015librispeech,lugosch2019alignments_zenodo,mcauliffe2017montreal,gilkeyio_librispeech_alignments}. The proposed schema involves constructing an oral-context versus nasal-context supervision signal and applying a sampling strategy that suppresses speaker and phonetic confounds by creating positive pairing on the same vowel from the same speaker and vowel-restricted contrastive comparisons. With frozen-encoder VPD screening, the learned encoder is frozen and used as a feature extractor for VPD screening with lightweight classifiers, aggregating chunk-level probabilities to yield recording-level screening decisions (in-domain and out-of-domain) using a fixed decision threshold. This design separates representation learning from clinical classification, aiming to improve cross-domain robustness without adaptation to the target domain \citep{liu2025out}.

\section{Materials and Methods}

\subsection{Datasets and Preprocessing}

\subsubsection{Clinical In-Domain Cohort}
The in-domain cohort of 82 patients was collected under standardized acoustic conditions in a clinical setting. Clinical labels were assigned based on \emph{human} specialist speech-language pathology (SLP) perceptual speech evaluation as part of routine clinical assessment, supported by instrumental assessments such as videonasoendoscopy (VNE) and/or nasometry when indicated. The VPD case cohort comprised 44 patients with clinically significant VPD as determined by the treating SLP, whereas 38 control patients were identified as having adequate velopharyngeal function based on \emph{human} SLP evaluation. Cohort definition and cohort partitioning followed our established protocol for automated VPD detection and out-of-domain (OOD) validation \citep{alter2025support,liu2025out}. 
All recordings were resampled to 16{,}000~Hz and converted to mono. Each audio file was segmented into multiple non-overlapping 0.5-second chunks. For files shorter than 0.5~s (and residual segments shorter than 0.5~s), we applied repeating padding by tiling the available audio content until reaching the 0.5-second target length (instead of zero-padding) to better preserve short-utterance acoustic characteristics. Each 0.5-second segment was treated as a modeling unit for feature extraction and chunk-level inference.

\subsubsection{Out-of-Domain Cohort}
To simulate real-world deployment under domain shift, we relied upon an OOD test set constructed from publicly available Internet sources. OOD control-group recordings were sourced from the Centers for Disease Control and Prevention and the Eastern Ontario Health Unit, and OOD case/control labeling and dataset curation followed the same protocol described in a prior OOD validation study \citep{liu2025out}. The resulting OOD set contains 131 recordings (70 controls, 61 VPD cases). These public sources do not provide reliable speaker identifiers or comprehensive metadata, so each recording is treated as an independent evaluation unit and \emph{recording-level} performance is reported on the OOD set. Models trained on the in-domain cohort were evaluated directly on the OOD recordings \emph{without any retraining, fine-tuning, or calibration} to quantify robustness under domain shift, consistent with prior practice in this clinical context \citep{liu2025out}. The same procedure used in the clinical in-domain dataset to sample and segment recordings was used for the OOD dataset.

\subsubsection{Nasality Pre-Training Corpus}
To learn a device- and content-robust representation of nasality prior to clinical classification, auxiliary pre-training on the \textit{Librispeech Alignments} dataset was performed \citep{gilkeyio_librispeech_alignments,lugosch2019alignments_zenodo}. This dataset is derived from the LibriSpeech corpus \citep{panayotov2015librispeech} and provides 16\,kHz read English speech with word- and phoneme-level alignments generated by the Montreal Forced Aligner (MFA) \citep{mcauliffe2017montreal}. This publicly available alignment resource was introduced in prior work on speech model pre-training for end-to-end spoken language understanding \citep{lugosch2019speech}. The alignments include phoneme boundaries, enabling time-localized extraction of vowel-centered segments. Using the phoneme-level alignment, we extracted short, vowel-centered acoustic segments from each utterance. Each segment is indexed by a \textit{(vowel, speaker)} key inferred from the filename and metadata. This key is relied upon to construct within-speaker and within-vowel comparisons to minimize confounds from speaker identity and vowel content.

An auxiliary binary supervision signal was constructed for contrastive learning using a rule-based labeling procedure derived from phoneme alignments. Using the MFA-provided ARPAbet phoneme boundaries in the LibriSpeech Alignments dataset, each vowel segment was extracted and assigned a context-based label using its immediate left and right neighboring consonants. Vowels flanked by two oral consonants (C--V--C, where C denotes a non-nasal consonant) were assigned to the \textbf{oral core} set, while vowels flanked by two nasal consonants (N--V--N; $N\in\{\text{M, N, NG}\}$) were assigned to the \textbf{nasal strong} set. Mixed contexts (C--V--N or N--V--C) were treated as \textbf{nasal weak} and excluded from SupCon training to reduce label noise from partial coarticulation. As basic quality control, we skipped utterances with missing audio bytes, required a 16~kHz sampling rate, and ignored vowels at utterance boundaries where a full left/right context is unavailable. The speaker identity relied upon for same-speaker sampling was inferred from the utterance ID prefix (e.g., \texttt{6415} in \texttt{6415-116629-0034}). These auxiliary labels are used \textit{only} for representation pre-training and are never used as clinical VPD diagnosis labels. All extracted segments were resampled to 16\,kHz, converted to mono, and standardized to a fixed duration of 0.20\,s (3200 samples) via center cropping (if longer) or zero padding (if shorter).

\subsection{Supervised Contrastive Learning for Nasality Representation}

\subsubsection{Pair Construction and Vowel-Restricted Supervised Contrastive Objective}
A nasality encoder was pre-trained using a supervised contrastive learning (SupCon) process \citep{khosla2020supervised}. This process extends the foundational self-supervised SimCLR framework \citep{chen2020simple} by leveraging label information to pull same-class samples together while pushing away opposite-class samples. A sampling strategy was employed specifically designed to suppress speaker and phonetic confounds. Each training item produces two ``views'' $(\mathbf{x}_1, \mathbf{x}_2)$ sampled from two \textit{different} segments belonging to the same class ($y\in\{0,1\}$), the same vowel, and the same speaker. A vowel `$v$' is sampled from the set of vowels that have at least two oral segments and at least two nasal segments. The class label `$y$' is then sampled with equal probability. Finally, we sample two distinct files from the corresponding bucket indexed by key $(v,\text{speaker})$:
\[
(\mathbf{x}_1,\mathbf{x}_2)\sim\mathcal{D}_{y,v,\text{speaker}}, \quad \mathbf{x}_1\neq \mathbf{x}_2.
\]

To further reduce phonetic content leakage, the contrastive objective is restricted to compare embeddings within the same vowel only. Cross-vowel pairs are excluded from the denominator in the contrastive softmax. In each training step, we sample a mini-batch of $B$ paired views, and the encoder outputs embeddings $\{\mathbf{z}_i^{(1)}, \mathbf{z}_i^{(2)}\}_{i=1}^{B}$, which we concatenate into a set of $2B$ $\ell_2$-normalized vectors. Let $y_i$ denote the oral/nasal label and $v_i$ the vowel ID. Because the embeddings are $\ell_2$-normalized, their dot product corresponds to cosine similarity, providing a scale-invariant measure of similarity commonly used in contrastive learning. Pairwise cosine similarities are computed and scaled by a temperature parameter $\tau$ to control the sharpness of the contrastive softmax distribution:
\[
s_{ij} = \frac{\mathbf{z}_i^\top \mathbf{z}_j}{\tau}, \quad \tau=0.07.
\]
Self-similarities are masked ($i=j$) and comparisons are restricted to samples with the same vowel ($v_i=v_j$). The positive set for anchor $i$ consists of all samples $j\neq i$ such that $y_j=y_i$ and $v_j=v_i$. The vowel-restricted SupCon loss is:
\[
\mathcal{L} = -\frac{1}{2B}\sum_{i=1}^{2B}\frac{1}{|\mathcal{P}(i)|}\sum_{p\in \mathcal{P}(i)}
\log\frac{\exp(s_{ip})}{\sum_{a\in \mathcal{A}(i)}\exp(s_{ia})},
\]
where $\mathcal{A}(i)=\{j\neq i: v_j=v_i\}$ and $\mathcal{P}(i)\subset\mathcal{A}(i)$.

\subsubsection{Encoder Architecture: Wav2Vec2 with Layer Fusion and Partial Unfreezing}
A Wav2Vec2-style transformer encoder is used as the backbone feature extractor \citep{baevski2020wav2vec}. The backbone is initialized from a Wav2Vec2-Large-960h pretrained checkpoint that is stored locally (``wav2vec2-large-960h-local''). The final $K = 4$ hidden layers are then fused with learnable weights, and the pooled representation is projected to a 256-dimensional embedding, where $\mathbf{H}^{(\ell)}\in\mathbb{R}^{B\times S\times d}$ denotes the hidden states from layer $\ell$. A weighted sum of the final four layers is then computed:
\[
\mathbf{H}_{\text{fused}} = \sum_{i=1}^{K} \alpha_i \mathbf{H}^{(L-K+i)},\quad
\alpha_i = \frac{\exp(w_i)}{\sum_{j=1}^{K}\exp(w_j)},
\]
where $\{w_i\}$ are learnable parameters and $L$ is the total number of transformer layers. Mean pooling is then applied over the sequence dimension:
\[
\mathbf{h} = \text{MeanPool}(\mathbf{H}_{\text{fused}})\in\mathbb{R}^{B\times d},
\]
followed by a two-layer MLP projection head to produce a 256-dimensional embedding:
\[
\mathbf{z} = \text{Normalize}\big(\text{BN}(\text{MLP}(\mathbf{h}))\big)\in\mathbb{R}^{B\times 256}.
\]
Embeddings are $\ell_2$-normalized. Batch normalization is applied during training when the batch size is greater than 1. To balance adaptation and stability, the backbone is frozen except for the last $N = 4$ transformer layers (and the encoder layer normalization), while always training the layer-fusion weights and the projection head. In summary, the first $L - 4$ transformer layers are kept fixed and only the last four layers are fine-tuned during SupCon pre-training.

\subsubsection{Training Details}
The model is then trained using AdamW with weight decay 0.01 and two parameter groups to implement layer-wise learning rates: (i) fusion/projection head parameters at $3\times 10^{-4}$, and (ii) unfrozen backbone layers at $3\times 10^{-5}$. Gradients were clipped to a maximum norm of 5.0, mixed precision training used bfloat16, and training ran up to 20 epochs with early stopping (patience = 6) based on the validation monitor below. A random 10\% split of the auxiliary dataset was used as validation (seed = 42). Batch size was auto-selected from \{2048, 4096, 6144, 8192\} to maximize throughput under GPU memory. Unless otherwise specified, all Wav2Vec2 backbone weights were initialized from the same Wav2Vec2-Large-960h pretrained checkpoint and loaded from a locally cached copy for training efficiency.

\begin{table}[h!]
\centering
\caption{Key hyperparameters for nasality supervised contrastive (SupCon) pre-training.}
\label{tab:supcon_hparams}
\begin{tabular}{ll}
\hline
Component & Setting \\
\hline
Backbone & Wav2Vec2-Large-960h (local checkpoint), last-layer fusion \\
Input sampling rate & 16 kHz \\
Segment duration & 0.20 s (3200 samples), center crop / zero pad \\
Embedding dimension & 256 \\
Layer fusion & last $K = 4$ layers, learnable softmax weights \\
Unfreezing & last $N = 4$ transformer layers + encoder layer norm \\
Loss & vowel-restricted supervised contrastive loss \\
Temperature & $\tau = 0.07$ \\
Optimizer & AdamW \\
Learning rates & head: $3\times 10^{-4}$; backbone: $3\times 10^{-5}$ \\
Weight decay & 0.01 \\
Gradient clipping & max norm 5.0 \\
Precision & AMP bfloat16 \\
Epochs / early stop & up to 20; patience 6 (monitor: validation pairwise accuracy; SupCon only) \\
Validation split & 10\% random split; seed = 42 \\
Batch size & auto-selected from \{2048,4096,6144,8192\} \\
\hline
\end{tabular}
\end{table}

\subsubsection{Representation Quality Monitor}
The embedding space was evaluated using a pairwise distance discrimination task on the validation split. Under matched vowel and matched speaker, positive pairs are same-class (oral--oral or nasal--nasal) and negative pairs are cross-class (oral--nasal). Given embeddings $\mathbf{z}_1,\mathbf{z}_2$, the Euclidean distance was computed:
\[
d = \lVert \mathbf{z}_1 - \mathbf{z}_2\rVert_2.
\]
Representation quality was evaluated using a distance-based pairwise discrimination task under matched speaker and matched vowel. For early stopping, validation pairwise accuracy is computed by selecting a fixed distance threshold on the validation split and applying it consistently across vowels. This monitor is used only to track representation learning during SupCon pre-training; it is not used for the downstream VPD screening classifier, which uses a fixed decision threshold on predicted probabilities.

\subsection{Lightweight VPD Classification}
After pre-training, the encoder is frozen and 256-dimensional embeddings are extracted for each 0.5-second chunk in the clinical datasets. Several lightweight classifiers are trained on top of these embeddings using only the in-domain training data: logistic regression (LR), support vector machine (SVM), multilayer perceptron (MLP), and XGBoost \citep{chen2016xgboost}. These models are computationally efficient and represent complementary decision functions (linear, margin-based, shallow nonlinear, and gradient-boosted trees), allowing a well-performing yet deployment-friendly classifier to be selected via cross-validation without changing the underlying representation. For all classifiers, a standardization step (z-score normalization) fit was applied on the training fold only. The hyperparameters were selected via group-wise cross-validation on the in-domain training split using \texttt{GroupKFold} to prevent subject leakage, where all recordings (and their constituent chunks) from the same subject were kept in the same fold. The selection criterion was chosen to be recording-level macro-F1 averaged across folds, where chunk-level probabilities were first aggregated within each recording. The searched grids were:
\begin{itemize}
    \item \textbf{LR:} $C \in \{0.01, 0.1, 1, 10\}$ (class-weight balanced).
    \item \textbf{SVM:} we tuned the kernel type and associated hyperparameters (e.g., $C$ for linear; $C$ and $\gamma$ for RBF) using cross-validation (class-weight balanced; probability outputs enabled).
    \item \textbf{MLP:} hidden sizes $\in \{(64), (128,64)\}$, activation $\in \{\text{relu}, \text{tanh}\}$, and $\alpha \in \{10^{-4}, 10^{-3}\}$ with early stopping.
    \item \textbf{XGBoost:} we tuned standard gradient-boosted tree hyperparameters (e.g., max depth, number of estimators, and learning rate) via cross-validation.
\end{itemize}
For brevity, the SVM results are reported without specifying the kernel in the tables. The best performing SVM configuration was selected by cross-validation. After setting the hyperparameters, each classifier was refit on the full in-domain training split and evaluated on the held-out test split.

\subsection{VPD Classification Baseline Models}
To ensure a fair and direct comparison with prior work, commonly used baseline feature extractors and classical classifiers are evaluated under the same preprocessing and evaluation protocol. Five feature extraction pipelines are implemented for each 0.5-second chunk:
\begin{itemize}
    \item \textbf{MFCC (baseline) \citep{davis1980comparison}:} 40 MFCC coefficients per chunk, mean-pooled over frames (40-d).
    \item \textbf{Wav2Vec2-Large-960h (frozen) \citep{baevski2020wav2vec}:} final-layer hidden states mean-pooled over time (1024-d).
    \item \textbf{HuBERT-Large (frozen) \citep{hsu2021hubert}:} final-layer hidden states mean-pooled (1024-d).
    \item \textbf{Data2Vec-Audio-Large (frozen) \citep{baevski2022data2vec}:} final-layer hidden states mean-pooled (1024-d).
    \item \textbf{Whisper-Large-v2 encoder (frozen) \citep{radford2023robust}:} encoder hidden states mean-pooled (1280-d).
\end{itemize}

Four classifiers were evaluated for each feature type: SVM, logistic regression, MLP, and XGBoost \citep{chen2016xgboost} (20 feature-classifier pipelines). For SVM, the kernel and associated hyperparameters were selected through cross-validation. For brevity, SVM results are reported without kernel specification.

\subsection{Model Evaluation}

Recording-level screening performance is reported for both the in-domain clinical cohort and the OOD Internet recordings. Each recording is segmented into non-overlapping 0.5-second chunks and the classifier outputs a chunk-level probability $\hat{p}_i$ for each chunk. For a recording with $N$ chunks, a recording-level probability is computed by mean aggregation:
\begin{equation}
\hat{p}_{rec} = \frac{1}{N}\sum_{i=1}^{N} \hat{p}_{i}.
\end{equation}
A fixed decision threshold $t = 0.5$ is then applied to obtain the binary screening decision. For the OOD Internet recordings, each recording is treated as an independent evaluation unit and the same aggregation and thresholding procedure is applied.

To prevent subject leakage (i.e., recordings from the same individual appearing in both the training and test sets), train/test splits are defined at the subject level such that no subject was included in both the training and testing sets. Hyperparameter selection was performed with group-wise cross-validation (\texttt{GroupKFold}), where all recordings (and their constituent chunks) from the same subject were assigned to the same fold. This design prevents the model from exploiting repeated recordings from the same individual across training and evaluation. It should be noted that this retrospective clinical cohort may exhibit demographic differences between cases and controls (e.g., age and sex distributions), which can act as potential confounders. For the OOD dataset, speaker identities are not provided and cannot be reliably inferred from the public sources. Therefore, speaker-disjoint evaluation cannot be enforced and instead, evaluation is performed directly at the recording level without any retraining, fine-tuning, or calibration.

All metrics are computed at the recording level, and accuracy, macro-precision, macro-recall, and macro-F1 are reported. For each class $c \in \{0,1\}$ (either control or VPD), precision, recall, and F1 are:
\begin{equation}
\mathrm{Prec}_c = \frac{\mathrm{TP}_c}{\mathrm{TP}_c+\mathrm{FP}_c},\quad
\mathrm{Rec}_c = \frac{\mathrm{TP}_c}{\mathrm{TP}_c+\mathrm{FN}_c},\quad
\mathrm{F1}_c = \frac{2\,\mathrm{Prec}_c\,\mathrm{Rec}_c}{\mathrm{Prec}_c+\mathrm{Rec}_c}.
\end{equation}
And macro-precision, macro-recall, and macro-F1 are computed as the unweighted mean across classes:
\begin{equation}
\mathrm{MacroPrec} = \frac{1}{2}\sum_{c}\mathrm{Prec}_c,\quad
\mathrm{MacroRec} = \frac{1}{2}\sum_{c}\mathrm{Rec}_c,\quad
\mathrm{MacroF1} = \frac{1}{2}\sum_{c}\mathrm{F1}_c.
\end{equation}
Accuracy is computed as the fraction of correctly classified recordings.

\section{Results}

\subsection{Dataset Summary}
\label{sec:dataset_stats}

The in-domain cohort consists of 82 subjects and was partitioned into an in-domain 60-subject training set and a 22-subject held-out test set using a subject-disjoint split to prevent subject leakage. The training set comprised 28 controls and 32 VPD cases with 345 recordings, while the held-out test set comprised 10 controls and 12 VPD cases with 131 recordings. In the 60-subject training cohort, the controls were 71.4\% female with a mean age of $29.6 \pm 11.8$ years, while the VPD cases were 46.9\% female with a mean age of $10.0 \pm 3.8$ years. In the 22-subject held-out test cohort, the controls were 70.0\% female with a mean age of $30.8 \pm 10.2$ years, while the VPD cases were 41.7\% female with a mean age of $9.1 \pm 3.6$ years. Each subject contributed one or more recordings; therefore, although the split is defined at the subject level, model performance is reported at the recording level.

The OOD set contains 131 recordings (70 controls, 61 VPD cases), collected under largely undocumented recording conditions (device, environment, and channel characteristics), thereby introducing substantial heterogeneity. Because these public sources do not provide reliable speaker identifiers or comprehensive metadata, we treat each recording as an independent evaluation unit and report recording-level performance on the OOD set. The dataset for nasality SupCon pre-training includes 778{,}110 oral\_core segments and 42{,}670 nasal\_strong segments, with 406{,}473 nasal\_weak segments set aside to reduce label noise.

\subsection{Nasality Representation Pre-Training Validation}
Evaluation is performed to determine whether the SupCon pre-training objective yields a meaningful nasality-focused embedding space under controlled comparisons (matched speaker and matched vowel). On the validation split, the nasality encoder achieved the highest validation pairwise accuracy at epoch 19. Using a fixed distance-based decision rule on the validation split to discriminate matched-speaker, matched-vowel positive vs.\ negative pairs, the pairwise validation accuracy reached 0.724. This validation monitor is used only for SupCon pre-training early stopping and is separate from downstream VPD screening evaluation.

\begin{figure}[t]
\centering
\includegraphics[width=0.75\linewidth]{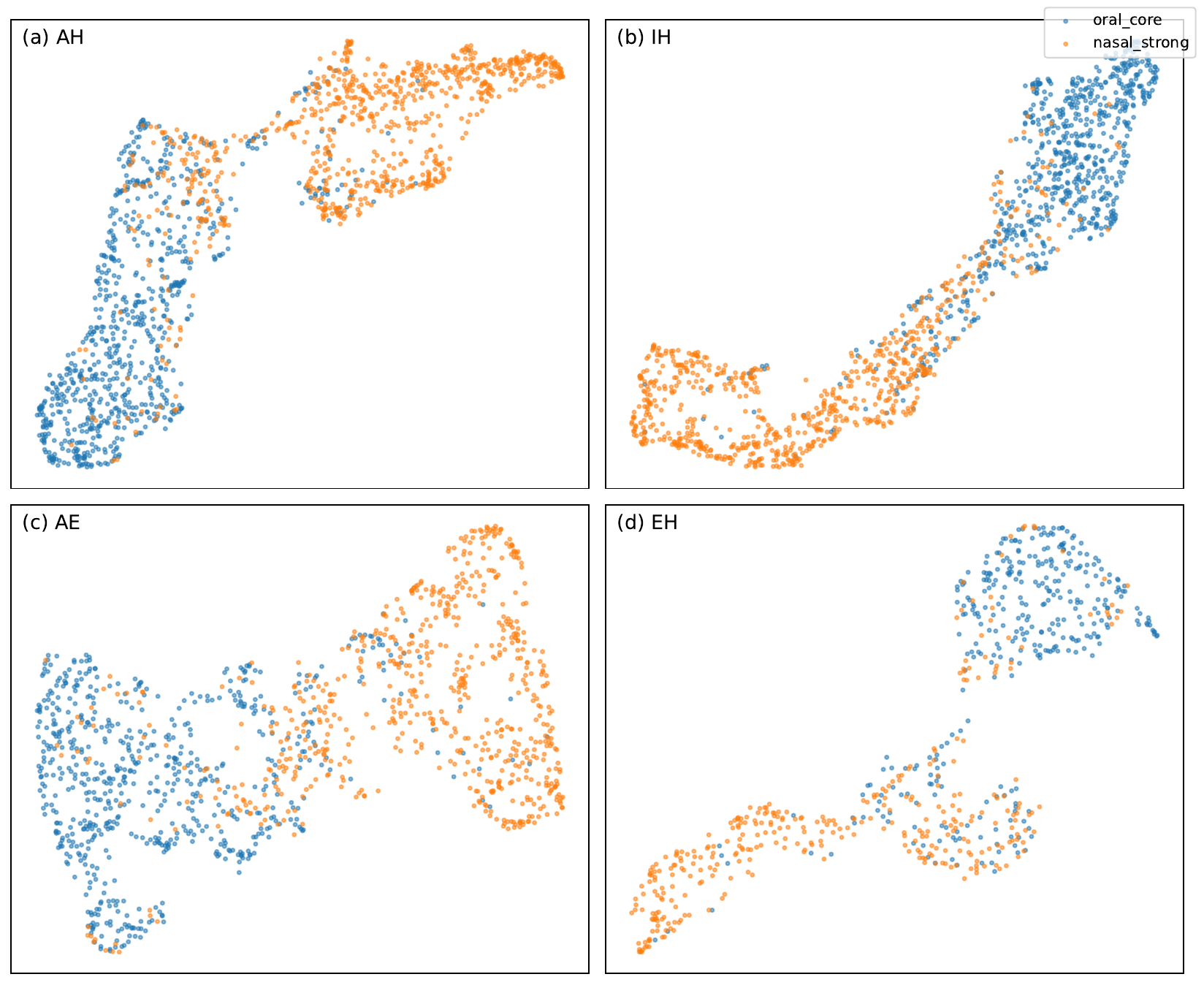}
\caption{
UMAP visualization of SupCon nasality embeddings on the auxiliary validation split.
Each panel shows a single vowel with a class-balanced subset of vowel-centered segments (0.20\,s).
Points are colored by the auxiliary nasality context label (oral\_core vs.\ nasal\_strong).
Vowel labels follow ARPAbet notation from the forced-alignment annotations.
}
\label{fig:umap_supcon_val}
\end{figure}

To qualitatively assess separation in the learned embedding space, validation embeddings are visualized using UMAP for multiple representative vowels (ARPAbet labels; e.g., AH, AE, IH, EH) with class-balanced subsets (\textbf{Fig.~\ref{fig:umap_supcon_val}}). Examination is performed to determine whether \textbf{oral core} and \textbf{nasal strong} segments form distinguishable clusters (or a consistent separation trend) within each vowel, indicating that the embedding captures nasality-related structure beyond vowel identity. Across the examined vowels, \textbf{oral core} and \textbf{nasal strong} segments show partial separation, consistent with the pairwise discrimination results.

\subsection{Screening Performance Under In-Domain and Out-of-Domain Evaluation}
VPD screening performance is evaluated using (i) baseline feature+classifier pipelines from the prior study evaluation protocol \citep{alter2025support,liu2025out} and (ii) the proposed SupCon nasality representation with lightweight classifiers. To streamline comparison and emphasize robustness, we report merged tables for in-domain and OOD evaluation (\textbf{Tables~\ref{tab:comparison_indomain_full} and~\ref{tab:comparison_ood_full}}).

For the in-domain held-out recordings, several pipelines achieved near-ceiling performance, i.e.\ mean performance at or near the upper bound observed on this standardized clinical dataset under the current evaluation protocol. Multiple large pretrained speech representations reached 100\% accuracy and macro-F1 of 1.000 (e.g., Whisper and HuBERT with MLP/XGBoost), indicating that under controlled recording conditions the screening task is highly separable for these data (\textbf{Table~\ref{tab:comparison_indomain_full}}). Using frozen 256-d SupCon nasality embeddings, all evaluated lightweight classifiers (LR/SVM/MLP/XGBoost) also achieved perfect recording-level screening performance (accuracy = 100\%, macro-F1 = 1.000). It should be noted that such perfect performance is specific to this in-domain cohort and protocol; it is observed across multiple strong baselines and is consistent with the high separability of the standardized clinical recordings and the recording-level aggregation used in evaluation, rather than implying uniformly perfect performance in more heterogeneous settings.

On the OOD dataset, performance drops substantially across most baseline pipelines, consistent with a strong domain shift between standardized clinical recordings and uncontrolled Internet audio (\textbf{Table~\ref{tab:comparison_ood_full}}). Among the baselines, MFCC+SVM is the strongest (macro-F1 = 0.612; accuracy = 64.1\%), while large pretrained speech representations degrade markedly (best macro-F1 among them = 0.432). The proposed SupCon nasality representation achieves the best overall OOD performance, with macro-F1 = 0.679 and accuracy = 69.5\% (MLP), improving over the strongest baseline by +0.067 macro-F1 and +5.4 accuracy points under the same evaluation protocol and fixed threshold.

\begin{table}[h!]
\centering
\scriptsize
\setlength{\tabcolsep}{3.5pt}
\renewcommand{\arraystretch}{1.08}
\caption{Recording-level performance on the \textbf{in-domain} held-out recordings (standardized clinical recordings; 131 recordings from 22 held-out subjects). This table merges (i) all baseline pipelines under the prior study protocol with (ii) the new SupCon nasality representation with lightweight classifiers. \textbf{Bold} highlights the proposed method for ease of comparison.}
\label{tab:comparison_indomain_full}
\begin{tabular}{llcccc}
\hline
Feature / Method & Classifier & Accuracy & Macro Prec. & Macro Rec. & Macro F1 \\
\hline
\multicolumn{6}{l}{\textit{Baselines (prior study protocol)}} \\
\hline
Whisper & MLP & 1.000 & 1.000 & 1.000 & 1.000 \\
Whisper & XGBoost & 1.000 & 1.000 & 1.000 & 1.000 \\
Whisper & SVM & 1.000 & 1.000 & 1.000 & 1.000 \\
Whisper & Logistic Regression & 1.000 & 1.000 & 1.000 & 1.000 \\
HuBERT & MLP & 1.000 & 1.000 & 1.000 & 1.000 \\
HuBERT & XGBoost & 1.000 & 1.000 & 1.000 & 1.000 \\
HuBERT & SVM & 0.992 & 0.929 & 0.996 & 0.960 \\
Data2Vec & XGBoost & 0.992 & 0.929 & 0.996 & 0.960 \\
MFCC & SVM & 0.992 & 0.996 & 0.917 & 0.953 \\
MFCC & MLP & 0.992 & 0.996 & 0.917 & 0.953 \\
MFCC & Logistic Regression & 0.992 & 0.996 & 0.917 & 0.953 \\
Data2Vec & Logistic Regression & 0.985 & 0.875 & 0.992 & 0.925 \\
HuBERT & Logistic Regression & 0.985 & 0.875 & 0.992 & 0.925 \\
MFCC & XGBoost & 0.985 & 0.913 & 0.913 & 0.913 \\
Data2Vec & SVM & 0.969 & 0.800 & 0.984 & 0.867 \\
Data2Vec & MLP & 0.954 & 0.750 & 0.976 & 0.821 \\
Wav2Vec2 & XGBoost & 0.946 & 0.731 & 0.972 & 0.801 \\
Wav2Vec2 & SVM & 0.946 & 0.731 & 0.972 & 0.801 \\
Wav2Vec2 & MLP & 0.930 & 0.700 & 0.963 & 0.767 \\
Wav2Vec2 & Logistic Regression & 0.907 & 0.667 & 0.951 & 0.724 \\
\hline
\multicolumn{6}{l}{\textit{Proposed (SupCon nasality representation; 256-d embeddings from 0.5\,s chunks)}} \\
\hline
SupCon Nasality (256-d) & Logistic Regression & \textbf{1.000} & \textbf{1.000} & \textbf{1.000} & \textbf{1.000} \\
SupCon Nasality (256-d) & SVM & \textbf{1.000} & \textbf{1.000} & \textbf{1.000} & \textbf{1.000} \\
SupCon Nasality (256-d) & MLP & \textbf{1.000} & \textbf{1.000} & \textbf{1.000} & \textbf{1.000} \\
SupCon Nasality (256-d) & XGBoost & \textbf{1.000} & \textbf{1.000} & \textbf{1.000} & \textbf{1.000} \\
\hline
\end{tabular}
\end{table}

\begin{table}[h!]
\centering
\scriptsize
\setlength{\tabcolsep}{3.5pt}
\renewcommand{\arraystretch}{1.08}
\caption{Recording-level performance on the \textbf{out-of-domain} test set (heterogeneous public Internet recordings). The table merges (i) all baseline pipelines under the prior study protocol and (ii) the proposed SupCon nasality representation with lightweight classifiers. All metrics use a fixed decision threshold of $t=0.5$.}
\label{tab:comparison_ood_full}
\begin{tabular}{llcccc}
\hline
Feature / Method & Classifier & Accuracy & Macro Prec. & Macro Rec. & Macro F1 \\
\hline
\multicolumn{6}{l}{\textit{Baselines (prior study protocol)}} \\
\hline
MFCC & SVM & 0.641 & \textbf{0.763} & 0.663 & 0.612 \\
MFCC & Logistic Regression & 0.603 & 0.745 & 0.628 & 0.560 \\
MFCC & XGBoost & 0.550 & 0.754 & 0.579 & 0.473 \\
MFCC & MLP & 0.512 & 0.619 & 0.540 & 0.432 \\
Wav2Vec2 & MLP & 0.512 & 0.619 & 0.540 & 0.432 \\
Wav2Vec2 & Logistic Regression & 0.504 & 0.588 & 0.532 & 0.427 \\
Data2Vec & SVM & 0.504 & 0.605 & 0.533 & 0.419 \\
Data2Vec & MLP & 0.489 & 0.537 & 0.515 & 0.417 \\
Data2Vec & Logistic Regression & 0.473 & 0.495 & 0.498 & 0.413 \\
Wav2Vec2 & SVM & 0.504 & 0.671 & 0.535 & 0.402 \\
Data2Vec & XGBoost & 0.481 & 0.527 & 0.509 & 0.397 \\
Wav2Vec2 & XGBoost & 0.489 & 0.571 & 0.518 & 0.393 \\
HuBERT & MLP & 0.504 & 0.742 & 0.536 & 0.393 \\
HuBERT & Logistic Regression & 0.489 & 0.738 & 0.521 & 0.364 \\
HuBERT & SVM & 0.481 & 0.736 & 0.514 & 0.349 \\
HuBERT & XGBoost & 0.481 & 0.736 & 0.514 & 0.349 \\
Whisper & Logistic Regression & 0.473 & 0.735 & 0.507 & 0.334 \\
Whisper & XGBoost & 0.473 & 0.735 & 0.507 & 0.334 \\
Whisper & MLP & 0.473 & 0.735 & 0.507 & 0.334 \\
Whisper & SVM & 0.466 & 0.233 & 0.500 & 0.318 \\
\hline
\multicolumn{6}{l}{\textit{Proposed (SupCon nasality representation; 256-d embeddings; recording-level aggregation)}} \\
\hline
SupCon Nasality (256-d) & MLP & \textbf{0.695} & 0.712 & \textbf{0.683} & \textbf{0.679} \\
SupCon Nasality (256-d) & SVM & 0.672 & 0.680 & 0.661 & 0.658 \\
SupCon Nasality (256-d) & Logistic Regression & 0.664 & 0.662 & 0.662 & 0.662 \\
SupCon Nasality (256-d) & XGBoost & 0.655 & 0.659 & 0.661 & 0.658 \\
\hline
\end{tabular}
\end{table}

\section{Discussion}
This investigation addresses a practical question in speech-based digital health: how can we maintain the performance of a model used to screen for speech pathology when moving from a regulated clinical setting into an unregulated field-testing environment? Motivated by the hypothesis that nasality-related cues may be more stable across recording conditions than many device- and environment-dependent artifacts, this study tested whether learning a nasality-focused representation prior to clinical classification improves robustness under recording domain shift. This hypothesis was evaluated by training models on the in-domain clinical cohort and assessing performance on a separate OOD Internet set without any target-domain retraining, fine-tuning, or calibration under a fixed screening threshold. Under this protocol, the proposed SupCon nasality representation achieved the best OOD performance (macro-F1 = 0.679, accuracy = 0.695), improving over the strongest retained baseline (MFCC+SVM: macro-F1 = 0.612, accuracy = 0.641). In contrast, in-domain performance was near-ceiling across multiple strong pipelines, underscoring that high in-domain accuracy does not guarantee deployable robustness and motivating domain-robust evaluation for VPD screening \citep{liu2025out}.

Four specific contributions should be noted. First, this workflow specifically studies VPD screening under substantial domain shift, from standardized clinical recordings to heterogeneous real-world/Internet recordings, using a unified evaluation protocol and a fixed screening threshold. Second, a supervised contrastive pre-training strategy is introduced that learns nasality-sensitive embeddings using oral-context versus nasal-context supervision derived from phoneme alignments, with sampling constraints to reduce speaker and phonetic-content leakage. Third, a two-stage, deployment-friendly pipeline was developed, with a frozen encoder feature extractor, lightweight classifiers, and simple probability aggregation. Fourth, this study provides controlled comparisons against MFCC features and large pretrained speech representations under the same protocol, demonstrating improved robustness in OOD testing. Collectively, these contributions provide evidence consistent with the hypothesis that nasality-focused representation learning can improve robustness under recording-domain shift, and they suggest a practical pathway toward scalable VPD screening on consumer devices where recording conditions are inherently variable and uncontrolled.

\subsection{Representation Learning Under Domain Shift}
A possible explanation for domain shift is that models trained and tested only on standardized clinical recordings can leverage recording-condition cues that are stable within a given setting (e.g., channel response, compression characteristics, background noise profiles, or room acoustics) but do not reflect pathology. As a result, performance can appear near-ceiling in-domain yet degrade sharply when those recording conditions change in real-world audio \citep{liu2025out,ganin2016domain}. In our in-domain cohort, several strong feature+classifier pipelines achieve perfect recording-level performance under the current protocol, suggesting that the standardized clinical recordings are highly separable for this dataset and that recording-level aggregation further amplifies separability. Perfect in-domain testing results can therefore be interpreted as the upper bound for this specific cohort and evaluation setting; however, this performance cannot be implied across non-standardized acoustic settings.

SupCon pre-training provides a more direct mechanism to emphasize nasality-related structure. The oral-context versus nasal-context supervision, together with same-speaker/same-vowel positive pairing and vowel-restricted contrastive comparisons, encourages embeddings to group according to nasality context while suppressing speaker identity and phonetic-content confounds \citep{khosla2020supervised}. This focus on nasality-related distinctions is consistent with the improved out-of-domain screening performance observed without any target-domain adaptation \citep{liu2025out}. Future work will aim to mitigate potential recording-condition shortcuts in the in-domain cohort by quantifying recording quality and channel artifacts (e.g., SNR/noise level, reverberation proxies, codec/compression indicators). Additionally, performing sensitivity analyses via stratification or exclusion of low-quality recordings may provide a clearer estimate of in-domain performance under varying acoustic conditions.

\subsection{Implications for Digital Health Deployment}
The pipeline presented in this study is referred to as the ``nasality pretrained screening (NPS)'' pipeline, which is deployment friendly given the frozen encoder and lightweight downstream classifier. This design supports real-world screening workflows, where recordings may be captured using diverse consumer devices and acoustic environments (e.g., home or web-sourced recordings), and inference can be performed either on device or in a server-assisted manner depending on resource constraints. In addition, recording-level aggregation by averaging chunk probabilities provides a simple and interpretable decision mechanism that can be used consistently across settings, supporting a screening workflow where repeated measurements may be collected longitudinally. This deployment-oriented framing is aligned with broader efforts in voice-based digital biomarkers and scalable screening pipelines beyond VPD \citep{liu2025voice}.

\subsection{Limitations}
This study has several limitations that should be acknowledged. First, the nasality pre-training supervision is derived from phoneme-alignment context rules (oral vs.\ nasal neighboring consonants). Because this rule-based labeling provides only an approximate proxy for nasality and may be noisy or incomplete, the resulting pretrained representation may not capture all clinically relevant variability in co-articulation, speaking style, and severity-dependent acoustic patterns. Second, the OOD dataset may contain uncontrolled confounders (device, environment, demographics, compression, and recording protocols), and the heterogeneous public Internet sources provide limited metadata, which limits our ability to derive causal attribution of the performance changes to any single factor \citep{liu2025out}. However, the extreme variability of the OOD cohort likely exaggerates what would be encountered in the field, which may under-estimate model performance. Additionally, since reliable speaker identifiers are unavailable in the public sources, we treat each recording as an independent evaluation unit. Yet this recording-level assumption may be violated if some recordings are correlated (e.g., multiple clips from the same speaker). As such, future investigations should validate robustness under speaker-disjoint OOD designs when identifiers are available. Third, the in-domain clinical cohort is relatively small (only 22 subjects in the held-out test split) and the near-perfect performance under standardized conditions likely overestimate robustness in broader populations, different elicitation prompts, or alternative clinical workflows. Moreover, the in-domain case and control groups may differ in demographic characteristics (e.g., age distributions), which could be a significant confounding factor and should be addressed in larger matched cohorts. Finally, a fixed screening threshold ($t = 0.5$) was used for simplicity and comparability across methods. In practice, the operating point may need to be tuned to clinical priorities (e.g., prioritizing sensitivity for screening) and probability calibration may be required across deployment settings.

\subsection{Future Directions}
Next steps include collecting multi-device datasets with standardized clinical metadata to better characterize which recording-condition factors (microphone response, room acoustics, noise, and compression) drive generalization failures, and to support principled robustness evaluations under controlled perturbations. Additionally, the screening task will be extended beyond binary detection toward severity grading and longitudinal monitoring of treatment response. Methodologically, exploring domain-robust objectives (e.g., domain-adversarial learning or correlation-alignment style regularization) may further reduce sensitivity to recording artifacts \citep{ganin2016domain,sun2016deep}. Finally, motivated by deployment in resource-limited and multilingual settings, we aim to study cross-lingual generalization and multilingual VPD screening, with the goal of developing models that maintain strong diagnostic performance across languages such as English and Spanish without requiring extensive language-specific re-collection or re-training.

\section{Conclusion}
This study presents a nasality-focused representation learning framework to improve the robustness of model-based VPD screening under substantial recording-domain shift. Using supervised contrastive pre-training with oral-context versus nasal-context supervision, the proposed encoder learns nasality-relevant cues while suppressing speaker- and phonetic-content confounds, enabling a frozen feature extractor with lightweight downstream classifiers for recording-level screening. Under standardized in-domain clinical recording conditions, performance was near ceiling on the subject-disjoint held-out test set. More importantly for deployment, the proposed approach improved OOD screening performance on heterogeneous public Internet recordings without retraining or calibration, outperforming the strongest retained baseline under the same evaluation protocol and fixed threshold. Overall, these results suggest that explicitly targeting physiologically meaningful attributes at the representation level can reduce reliance on recording artifacts and support deployable speech-based screening in real-world settings. Future work will prioritize prospective multi-device data collection with standardized metadata and deployment-aware calibration to establish clinically robust operating points.

\section*{Conflict of Interest}
The authors declare that the research was conducted in the absence of any commercial or financial relationships that could be construed as a potential conflict of interest.

\section*{Funding}
This work was supported by the Department of Plastic Surgery at Vanderbilt University Medical Center.

\section*{Author Contributions}
\textbf{Author Contributions (CRediT):} Conceptualization: W.L., B.M., Z.Y., M.E.P.; Methodology: W.L., B.Q., B.M., Z.Y., M.E.P.; Software: W.L., B.Q.; Data curation: W.L., A.S., M.P., S.D., S.B., I.G., M.G.; Data preprocessing: W.L., A.S., M.P., S.D., S.B., I.G., M.G.; Formal analysis: W.L.; Validation: W.L., A.S., M.P., S.D., S.B., I.G., M.G.; Investigation (clinical data support): A.S., M.P., S.D., S.B., I.G., M.G., M.E.P.; Resources: M.E.P.; Visualization: W.L., B.Q.; Writing---original draft: W.L.; Writing---review \& editing: all authors; Supervision: B.M., Z.Y., M.E.P.; Project administration: W.L., M.E.P.; Funding acquisition: M.E.P.

All authors read and approved the final manuscript.

\section*{Acknowledgments}
We thank the Department of Plastic Surgery at Vanderbilt University Medical Center for support in clinical data collection and curation.

\section*{Data Availability}
The in-domain clinical recordings were collected and curated through the Department of Plastic Surgery at Vanderbilt University Medical Center. Due to privacy and ethical restrictions, these clinical data are not publicly available. Requests for access may be considered upon reasonable request to the corresponding author and subject to applicable institutional approvals and data use requirements.

The out-of-domain test audio samples were obtained from publicly available Internet sources; control samples were obtained from the Centers for Disease Control and Prevention and the Eastern Ontario Health Unit.

\section*{Ethics Statement}
This study was conducted after institutional review board approval (IRB No.\ 212135) at Monroe Carell Jr.\ Children's Hospital at Vanderbilt. A retrospective review of data from the Vanderbilt Voice Center was performed.

\bibliographystyle{plainnat}
\bibliography{references}

\end{document}